\documentclass[showpacs,preprint,amsmath,superscriptaddress,prl]{revtex4}  
\usepackage[dvips]{graphicx}
\usepackage{epsfig}
\usepackage{color}
\usepackage{subfigure}
\usepackage{soul}

\begin{document} 
\title{
Finite-size scaling of human-population distributions over fixed-size cells
and its relation to fractal spatial structure  
} 
\author{\'Alvaro Corral}
\email{alvaro.corral@uab.es} 
\affiliation{Centre de Recerca Matem\`atica,
Edifici C, Campus Bellaterra,
E-08193 Barcelona, Spain.
} 
\affiliation{Departament de Matem\`atiques,
Facultat de Ci\`encies,
Universitat Aut\`onoma de Barcelona,
E-08193 Barcelona, Spain}
\affiliation{Barcelona Graduate School of Mathematics, 
Edifici C, Campus Bellaterra,
E-08193 Barcelona, Spain
}\affiliation{Complexity Science Hub Vienna,
Josefst\"adter Stra$\beta$e 39,
1080 Vienna,
Austria
}
\author{Montserrat Garc\'{\i}a del Muro}
\affiliation{%
Departament de F\'{\i}sica de la Mat\`eria Condensada, 
Universitat de Barcelona, Mart\'{\i} i Franqu\`es 1, E-08028 Barcelona, Spain
}\affiliation{%
IN2UB, Universitat de Barcelona, Mart\'{\i} i Franqu\`es 1, E-08028 Barcelona, Spain
}

\begin{abstract} 
%
%
%
%

Using demographic data of high spatial resolution for a region in the south of Europe, we study the population over fixed-size spatial cells. 
We find that, counterintuitively, the distribution of the number of inhabitants per cell increases its variability when the size of the cells is increased. 
Nevertheless, the shape of the distributions is kept constant, which allows us to introduce a scaling law, analogous to finite-size scaling, with a scaling function reasonably well fitted by a gamma distribution.
This means that the distribution of the number of inhabitants per cell is stable or invariant under addition with neighboring cells (plus rescaling), defying the central-limit theorem, 
due to the obvious dependence of the random variables. 
The finite-size scaling implies a power-law relations between the moments of the distribution and its scale parameter, which are found to be related with the fractal properties of the spatial pattern formed by the population.
The match between theoretical predictions and empirical results is reasonably good.



  

\end{abstract} 

\maketitle


\section{Introduction}


Problems related to human population are going to be among the most pressing ones
our societies will face in the near future. 
The so-called new science of cities is trying to bring a holistic and cross-disciplinary 
perspective to analyze all sort of patterns and phenomena 
that appear in
large human aggregations
\cite{West_book,Barthelemy_cities_review,Arcaute_Ramasco,Rybski_Gonzalez}.
A fundamental quantity in this endeavor,
which has been studied since long ago,
is 
the ``size'' of human aggregations 
(villages, towns, cities, megalopolis) 
measured in terms of number of inhabitants.
This has been found to be, obviously, very broadly distributed,
ranging from dozens of individuals to several millions.
The Zipf's paradigm proposes that, for a given country or region, 
the distribution of the number of inhabitants in these aggregations 
(defined, for instance, as municipalities),
follows, at least for the largest values, 
a power-law distribution with an exponent of the probability density close to two \cite{Zipf_1949,Krugman,Cottineau,footnote_corral_garcia_del_muro2,Corral_Cancho}.
But this has been challenged by other authors, 
who suggest that a lognormal distribution provides a better fit
to the empirical data
than that of the power law \cite{Eeckhout,Levy_comment,Malevergne_Sornette_umpu}.

Leaving aside the adequacy of each distribution to describe the population of human aggregations, 
a drawback of this approach is that it does not take into account
the spatial degrees of freedom, i.e., how the individuals are distributed in the territory 
-- the required data 
is just a list of municipalities with the number of inhabitants in each one.
Recently, Ref. \cite{Corral_Arcaute} took a different point of view, 
using the exact spatial coordinates of the living place of all the individuals in 
a territory to define clusters of spatially connected individuals (or, more precisely, clusters of close living places). 
This provided a somewhat ``natural'' definition of what a human aggregation
(roughly speaking, a city) is (see also Refs. \cite{Rozenfeld,Jiang2011}), 
and allowed an unambiguous calculation of its number of inhabitants 
(dependent only on the width of the cells
that constitute the clusters or on the distance that discriminate if two individuals are connected or not). 


The resulting distribution of number of inhabitants of these clusters \cite{Corral_Arcaute}
turned out to be even broader than in the traditional approach,
with many clusters consisting only of one or two individuals,
but it was also found that the lognormal distribution (truncated from below) 
led to a more ``complete'' fit
than the power law, providing a good fit from about 10 individuals to the population
of the largest cluster \cite{footnote_corral_garcia_del_muro3,Corral_Gonzalez,Serra_Corral_Zipf}.
Reference \cite{Corral_Arcaute} also showed how this broadness
emerged through the integration of neighboring highly-populated cells,
highlighting the importance of spatial correlations in the distribution of individuals
through the territory.
In other words, destroying the spatial correlations (through reshuffling of cells)
destroys or reduces the broadness of the distribution of number of inhabitants.

The high-resolution human-population data used in Ref. \cite{Corral_Arcaute}
allows for an in-depth study of the spatial structure of human populations,
and this is what we undertake in the present paper.
This sort of questions have been widely addressed in ecology, 
regarding both plants or animals,
as spatial patterns play an important role in the spread of diseases, predation, mating, etc.
\cite{Borregard}.
When considering a (relatively) large region, 
one typically divides it into smaller cells of the same area, 
and counts the number of individuals of a certain species in each cell.
The simplest probability description of this random variable is given 
by the Poisson distribution, which assumes a random structureless population.

Nevertheless, individuals interact, between them and with other species.
In the cooperative case, it is expected that individuals will show a tendency
towards clustering in space, giving rise to clumped or patchy patterns,
with an index of dispersion (variance divided by mean, also called Fano factor) larger than one.
This has been described by diverse probability distributions \cite{Borregard,Zillio},
but more prominently by the negative binomial, 
which can be theoretically justified by a mixture of Poisson distributions whose rates
follow a gamma distribution (i.e., in each cell, the number of individuals is Poisson, 
but with different rates for different cells).
On the contrary, if competition dominates, one expects more even or regular spatial patterns, with an index of dispersion smaller than one.

Let us note that both the Poisson and the negative binomial distributions lead to spatial patterns that are scale dependent, in the sense that the shape parameters of both distributions change with the size of the cells: 
It is easy to show that if one doubles the area of the cells, 
the Poisson parameter also doubles, and this changes the shape of the Poisson distribution, which becomes sharper. 
For large cells one gets something very close to a Gaussian distribution,
and in the limit of very large cells one gets a Dirac delta (at the scale of the mean of the variable).
This property of the change of the shape parameter is inherited by the negative binomial, 
as this distribution is a mixture of Poisson distributions (we develop this in Appendix I).

In this paper, we simply count human individuals in small spatial cells, 
and compute the corresponding probability distributions.
Remarkably, we find that the distributions for different cell sizes are related through finite-size scaling, representing a ``stability'' of the distributions under cell aggregation (plus proper rescaling).
The corresponding power-law relations for the moments of the distributions allow us to relate this scaling with multifractals.


\section{Data}


We analyze high-resolution data 
for the living places of all citizens in Catalonia, 
a region in NE Spain and whose capital and largest city is Barcelona.
Catalonia 
has an area of about 32,000 km$^2$
and a total population slightly above 7,500,000 inhabitants 
(this yields an average population density around 230 inhabitants per km$^2$
and clasifies Catalonia as a highly populated area).
This figures are very similar to those of Switzerland, for example.
Interestingly, the metropolitan area of Barcelona 
(in a spot comprising the municipality of l'Hospitalet and part of Barcelona) 
has the highest population density in Europe (at the scale of 1 km$^2$) \cite{Rae}.

\begin{figure}[ht]
\includegraphics[width=.50\columnwidth]{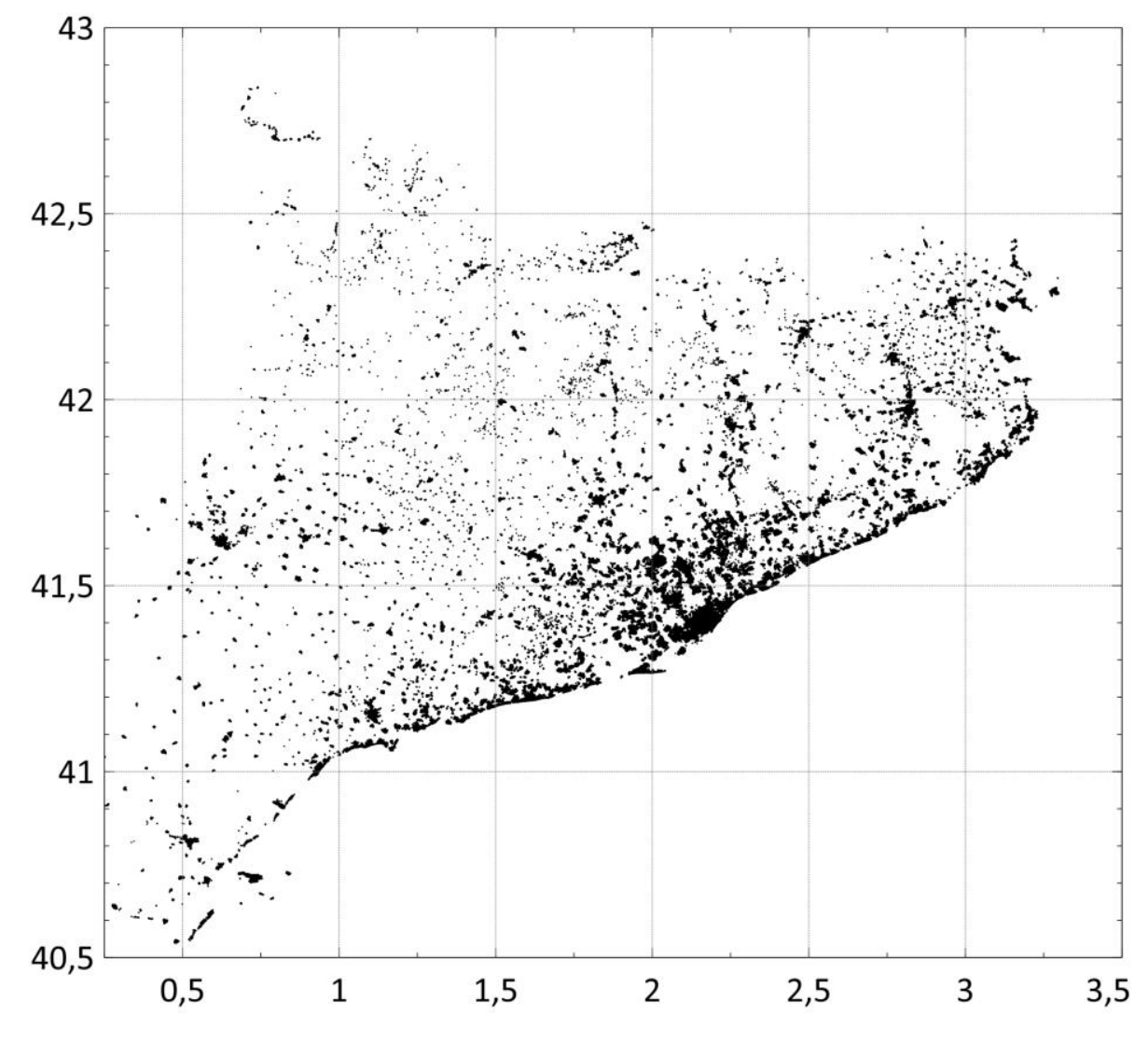}\\
\includegraphics[width=.50\columnwidth]{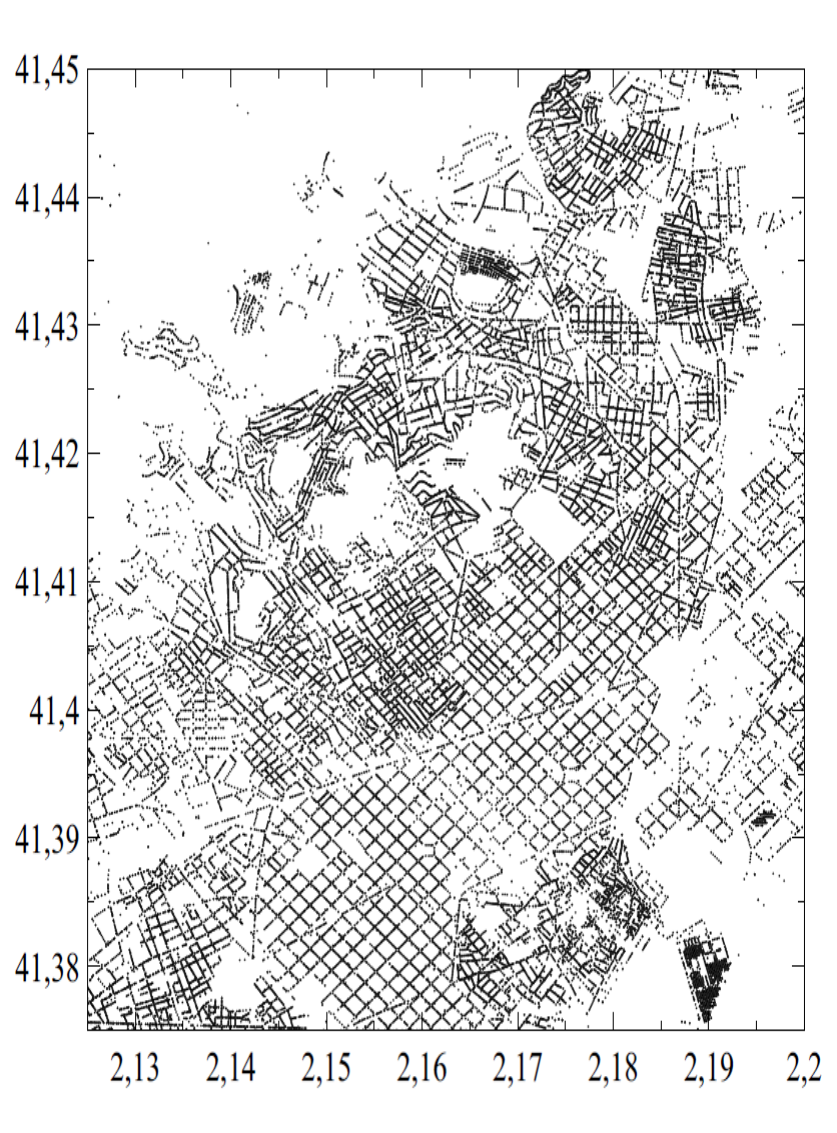}
\caption{
Top:
Latitude and longitude of the 7,586,888 inhabitans of Catalonia 
at January 1, 2013 (whole dataset).
Bottom:
Zoom of the data around the Barcelona area.
Notice that we are representing the coordinates of the residence place 
of each individual, so, discreteness effects become apparent
in the bottom plot.
}
\label{Catalonia}
\end{figure}

Each municipality council in Spain collects
a population register called {\it Padr\'on Municipal de Habitantes} 
and 
the {\it Institut d'Estad\'{\i}stica de Catalunya} (IDESCAT) 
receives the information referred to Catalonia.
IDESCAT georeferences
the postal address of each individual in the register, 
by means of the  
geocoding web service of the {\it Institut Cartogr\`afic i Geol\`ogic de Catalunya}.
The complete procedure 
is detailed in Ref. \cite{Sune}.
For this study we used the data
corresponding to the population of Catalonia
at January 1, 2013, 
which yields a total population of $M=$7,586,888 inhabitants
in 989,997 places of residence,
with a 7.6 \% of errors in the georeferencing,
for which an imputation procedure is applied \cite{Sune}.
This is the same data used in the study of Ref. \cite{Corral_Arcaute}.
The spatial distribution of the complete data set is displayed in Fig. \ref{Catalonia},
including also a zoom over the Barcelona area.
The spatial resolution of this data is higher than that of other previous studies using high-resolution data for Switzerland and France, for instance \cite{Orozco,Semecurbe}.


%

\section{Analysis}

We work in a simple equirectangular projection
of longitude and latitude into Cartesian coordinates;
this introduces little distortion due to the small
extent of the territory 
(Reference \cite{Corral_Arcaute} showed that the results there did not depend 
on the choice of the projection). 
The resulting projection is covered by a grid composed by 
identical square cells, each of size $\ell \times \ell$ (in degrees),
aligned with the longitude-latitude axes.
The range of values of $\ell$ that is of our interest is between $0.0002^\circ$ and $0.1^\circ$ 
(from 20 m to 10 km, roughly).
The minimum values of longitude and latitude in the data define the leftmost and bottom
coordinates of the grid, respectively \cite{footnote_corral_garcia_del_muro}.


\subsection{Occupation of cells}

The most fundamental issue in our approach 
is counting the number of inhabitants (individuals) $h$ in each cell.
For the values of cell width $\ell$ considered,
the resulting values of $h$ turn out to be broadly distributed,
from one inhabitant per cell to hundreds of thousands 
(we will disregard unpopulated cells, 
for reasons that will become clear later). 
Notice that $h$ is not as broadly distributed as the population of the clusters
of Ref. \cite{Corral_Arcaute}, reaching millions, 
as one cluster there can consist of many cells.

\begin{figure}[ht]
\includegraphics[width=.490\columnwidth]{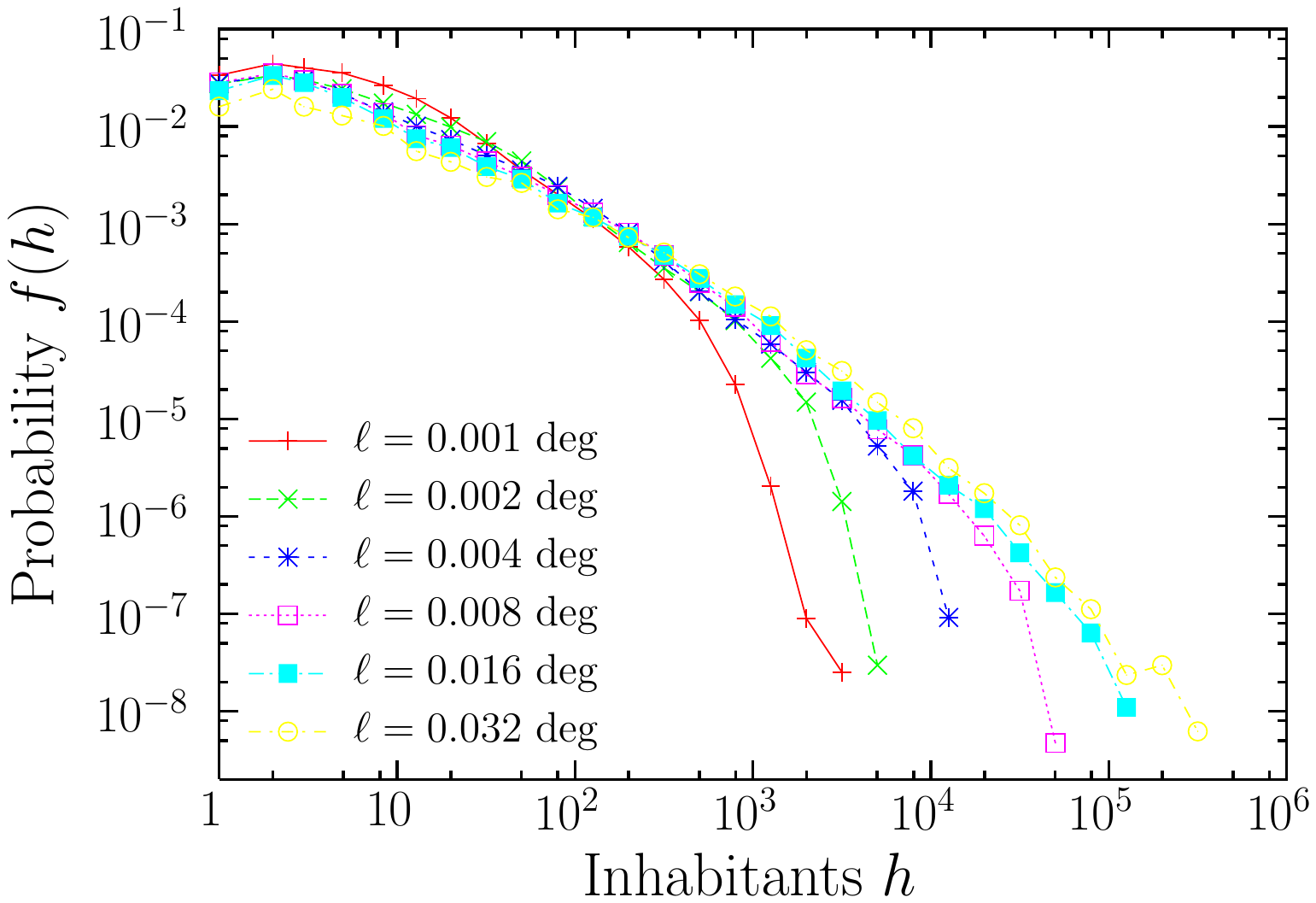}
\includegraphics[width=.490\columnwidth]{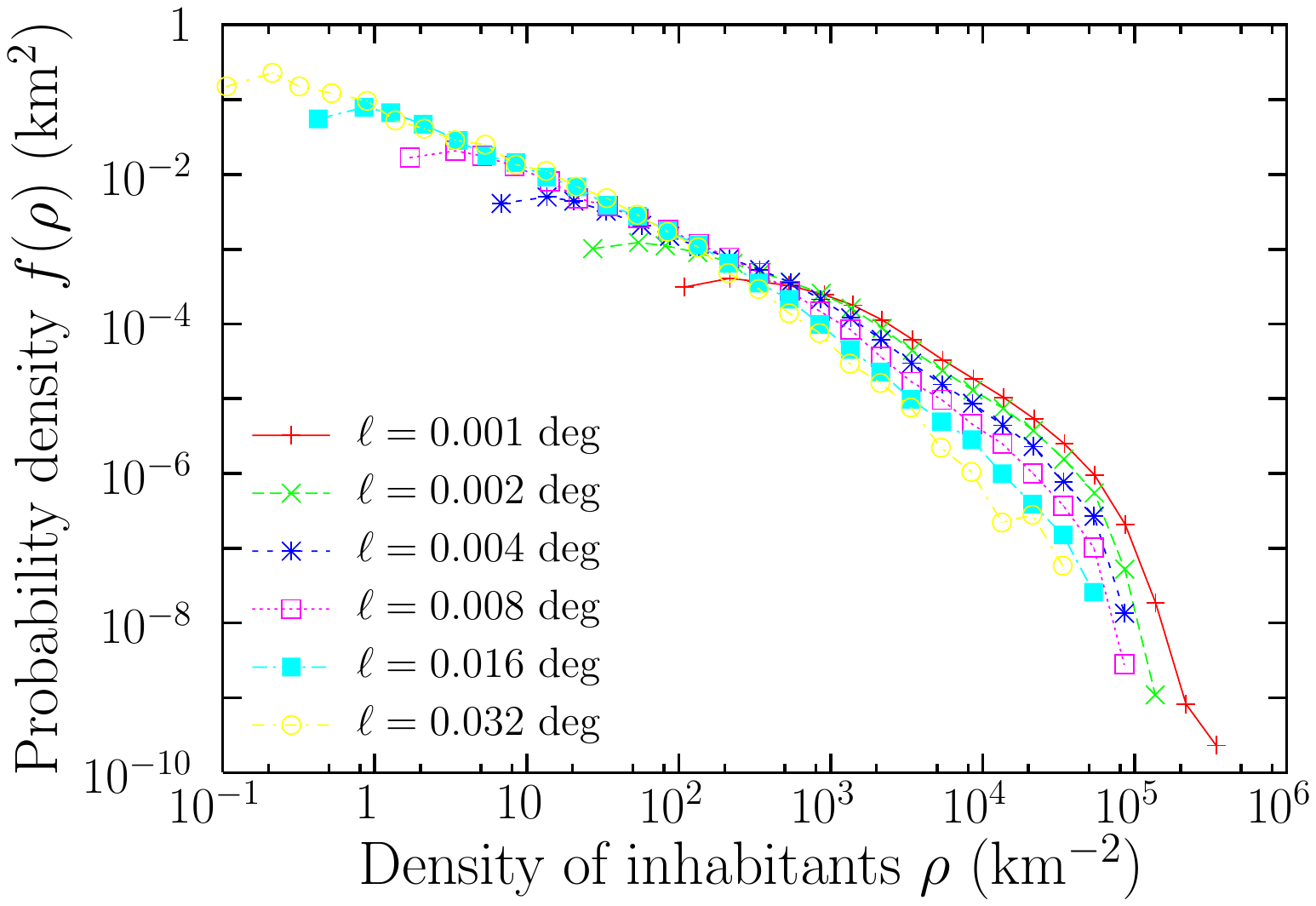}
\caption{(a) Empirical probability mass functions $f(h)$
of number of inhabitants $h$ per cell, 
for several values of cell width $\ell$. 
(b) Corresponding empirical probability densities $f(\rho)$ 
of population density $\rho$ per cell,
in units of km$^2$
(using that $1$ degree$^2 \simeq 9200$ km$^2$ at the latitude of Catalonia).
We find population densities far beyond $10^5$ inhabitants per km$^2$
(larger than the value reported in Ref. \cite{Rae}, as the size of our cells can be smaller).
}
\label{Dindividualcells}
\end{figure}

The corresponding probability mass function of the number of inhabitants, $f(h)$,
was shown in Ref. \cite{Corral_Arcaute};
nevertheless, for completeness, we show it also in Fig.~\ref{Dindividualcells}(a) 
for different values of $\ell$
(it would have been more precise to write $f_\ell(h)$ for $f(h)$,
but at this point we prefer to keep the notation at minimum).
As it can be seen in the figure,
the larger the value of $\ell$, the larger the variability in $h$;
this is counterintuitive, as under aggregation one naively expects that
fluctuations compensate and eventually become irrelevant, see Appendix I.

Further, 
the population density in each cell is calculated simply as $\rho=h/\ell^2$, 
and its probability density $f(\rho)$ is shown in Fig.~\ref{Dindividualcells}(b), 
for the sake of illustration.
The upper tails of these distributions contain the highest population
density in Europe, being reached next to Barcelona \cite{Rae}. 
The large variability in population densities 
implies that the concept of mean population density
is of little use for the description of population data
(being, in addition, scale dependent if unpopulated cells are disregarded, 
as we will see below).

\begin{figure}[ht]
\includegraphics[width=.490\columnwidth]{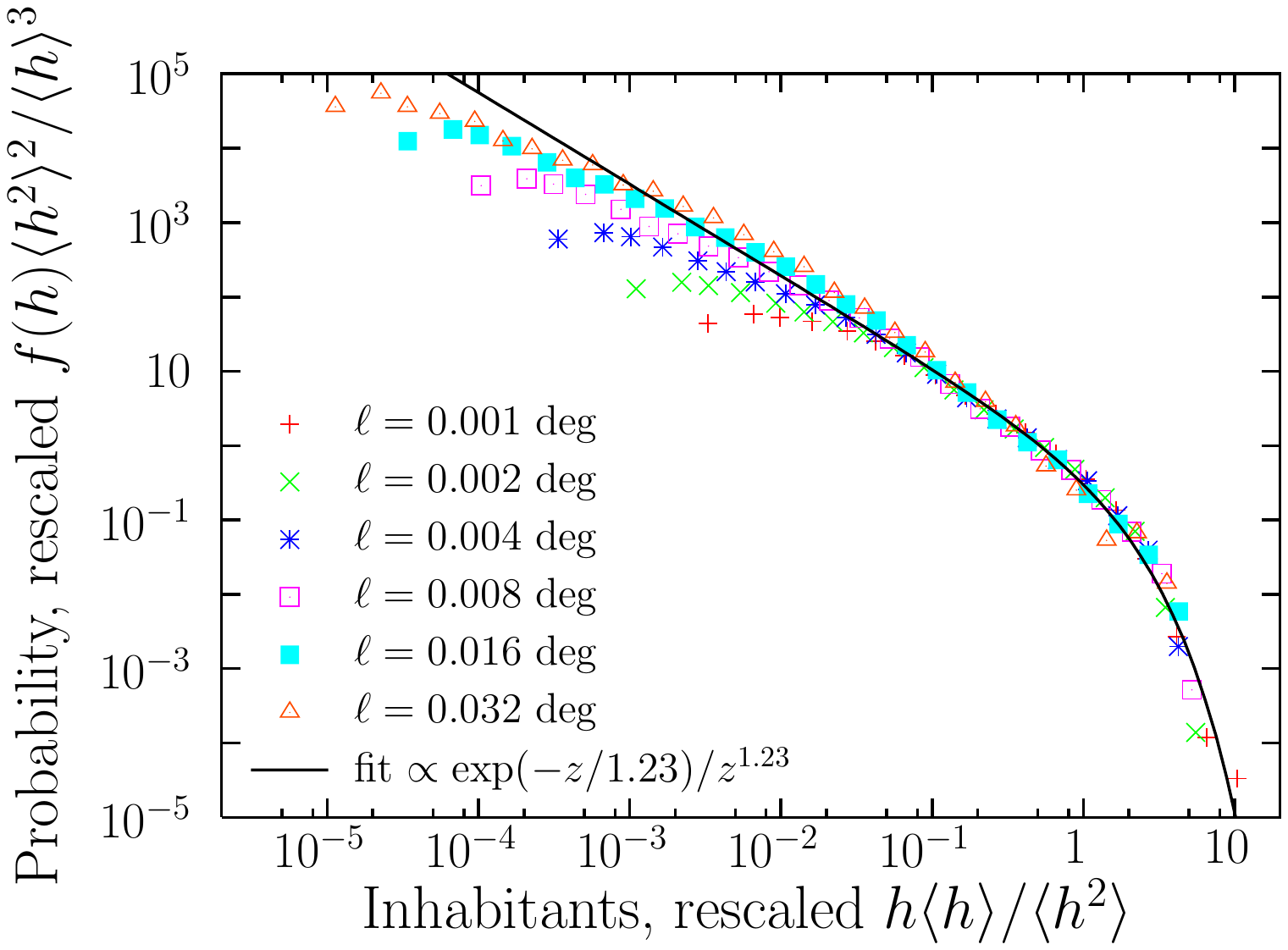}
\includegraphics[width=.490\columnwidth]{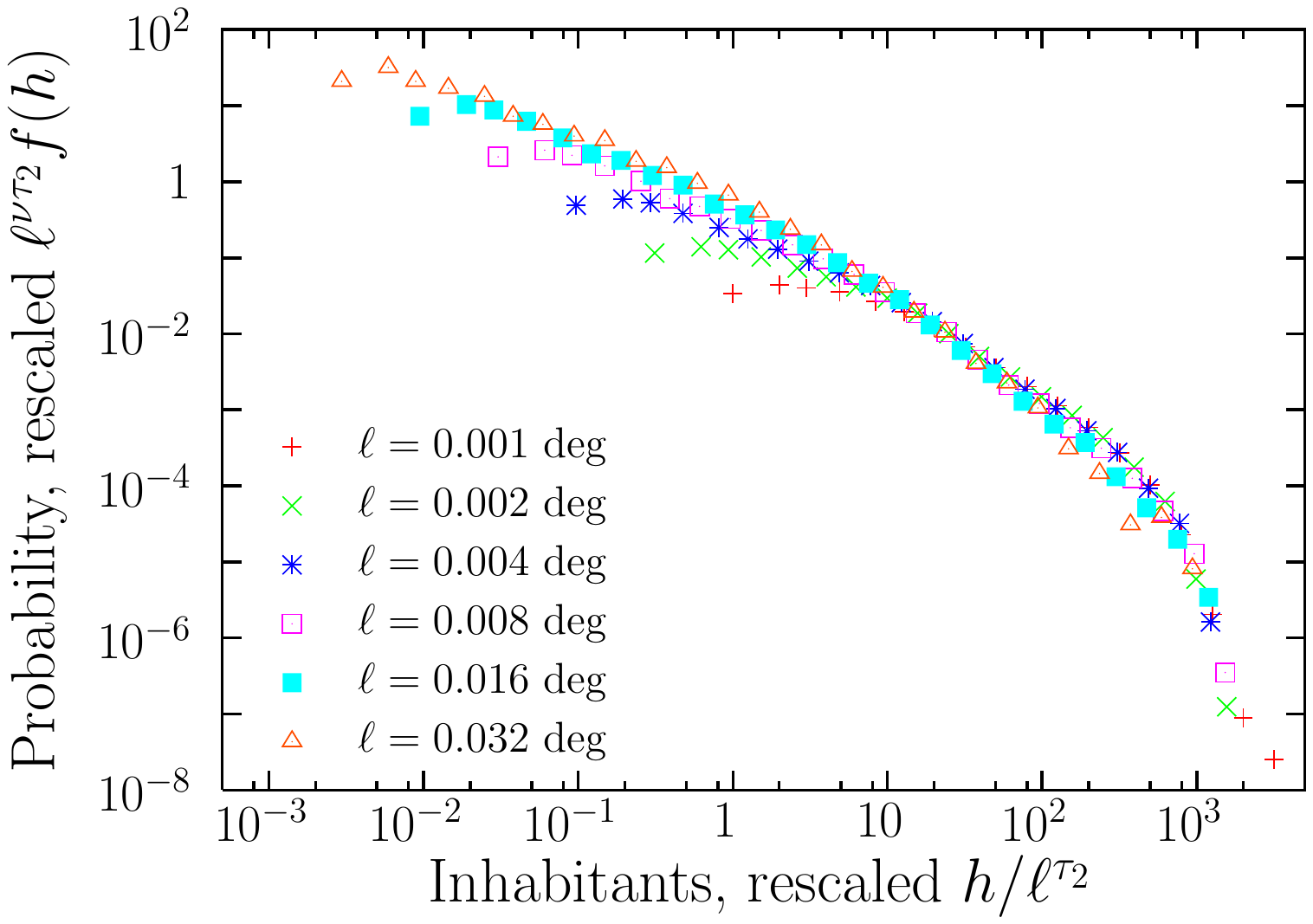}
\caption{
(a) Same distributions rescaled in terms of the moment ratios as 
stated in Eqs. (\ref{rescaling1}) and (\ref{rescaling2}).
The continuous line is a fit 
of the form $\propto e^{-z/\theta} /z^{1.23}$
in the range $z>0.05$
for all the curves,
resulting in $\theta=1.23$ 
(notice that $\nu$ is fixed to 1.23, as derived in the main text).
(b) Alternative scaling, corresponding to finite-size scaling, 
using $\tau_2=1.68$ and $\nu\tau_2=1.23\times1.68=2.07$.
In the rescaling, we have redefined $\ell$ as $\ell/1000$.
Both data collapses are an indication that all distributions have the same shape for $h >100$, roughly,
despite their different scales. 
}
\label{Dindividualcells2}
\end{figure}

\subsection{Scaling of the inhabitants-per-cell distribution}

A data collapse of all the distributions $f(h)$ 
for different cell width $\ell$ 
is possible (except for small values of $h$).
As Fig.~\ref{Dindividualcells2}(a) shows, 
and in contrast to the case of Ref. \cite{Corral_Arcaute},
this is not achieved by a naive scaling by the mean but 
by the rescaling 
\begin{equation}
h \rightarrow \frac{ \langle h\rangle}{\langle h^2\rangle}\, h,
\label{rescaling1}
\end{equation}
\begin{equation}
f(h) \rightarrow \frac{\langle h^2\rangle^2}{ \langle h\rangle^3}\, f(h),
\label{rescaling2}
\end{equation}
with $\langle h\rangle$ and $\langle h^2\rangle$
the (empirical) expected values of $h$ and $h^2$, respectively
(and equivalently for $\rho$ and $f(\rho)$;
notice that the rescaling makes the axes of the plot dimensionless),
see Ref. \cite{Corral_csf}.
The expected values should depend on $\ell$,
but we omit this dependence from the notation
(although it is capital).
The overlap of the rescaled distributions shown in  
the figure
indicates that all the distributions have the same shape for 
$z=h{ \langle h\rangle}/{\langle h^2\rangle} =\rho { \langle \rho\rangle}/{\langle \rho^2\rangle}  > 0.05$,
approximately.
So, in this range, 
it is only a scale parameter what distinguishes the distributions
for different $\ell$.


As it can be also seen in Fig.~\ref{Dindividualcells2}(a),
a gamma distribution truncated from below fits reasonably well the collapsed data.
Its probability density takes the form
\begin{equation}
f(z)\propto \left(\frac 1 {z}\right)^{1-\delta} e^{-z/\theta}
\label{gamma}
\end{equation}
with 
parameters $\delta$ and $\theta$.
The shape parameter $\delta$ turns out to be smaller than zero
(we will see below that $1-\delta\simeq 1.2$);
this implies that the standard deviation is larger than the mean
(at least in certain limit, see the scaling of moments below).
From the collapsed distributions displayed in the figure
it is clear that,
although the body of the distribution shows a power-law decay,
at the end, for the largest values of $h$,
the power law transforms into an exponential-like tail.



We will show in the next subsection that
the scaling property of $f(h)$ allows one to anticipate a fractal
behavior for the spatial distribution of the population.
First,
let us consider a general scaling law for the probability mass function of the
number of inhabitants per cell,
\begin{equation}
f(h) = \frac 1 m \left(\frac m h \right)^\nu G\left(\frac h c\right),
\label{threebis}
\end{equation}
valid for $h$ above a cut-off value $m>0$
(so $h=0$ is not counted, it is not considered an event), 
with $c$ a scale parameter, 
$\nu > 1$ a power-law exponent (being the shape parameter of the distribution), 
and $G$ a scaling function
that can be a decreasing exponential 
[corresponding to the case given by Eq. (\ref{gamma})]
or any other function with similar asymptotic properties
(going to a constant for small arguments and decaying fast for large ones).
Comparison with Eq. (\ref{gamma}) implies $\nu=1-\delta$, 
but $c\ne \theta$, as Eq. (\ref{gamma}) refers to the rescaled variable 
$z$ and the scaling law is written in terms of the number of inhabitants $h$.
%
We expect $c$ to increase with $\ell$,
whereas $\theta$ is a constant.
Note that the $\ell-$dependence of $f(h)$ is through the scale parameter $c$.
Below we will see that the scaling law (\ref{threebis})
is related with finite-size scaling.

The scaling property (\ref{threebis}) 
implies that the moments of the distribution 
scale (when $\nu > 1$), as
\begin{equation}
\langle h^q \rangle \propto 
\left\{
\begin{array}{ll}
{m^{q}} &\mbox{ for } q < \nu -1,\\
{m^{\nu-1}} {c^{q+1-\nu}} &\mbox{ for } q > \nu -1,\\
\end{array}
\right.
\label{ha2}
\end{equation}
see Ref. \cite{Christensen_Moloney} for $q > \nu -1$
or our Appendix II, in general.
The idea is that in the limit $c\rightarrow \infty$
all the moments above $\nu-1$ diverge, 
but those below do not.
From the previous expressions we can easily justify the rescaling in Eqs. (\ref{rescaling1})-(\ref{rescaling2});
indeed, for $\nu < 2$ consider $\langle h \rangle$
and $\langle h^2 \rangle$ to obtain the moment ratio
\begin{equation}
\frac{\langle h^2 \rangle }{\langle h \rangle } \propto c,
\label{momentratio}
\end{equation}
which is a particularly useful relation,  
see Refs. \cite{Corral_csf,Peters_Deluca}.
Further, one can notice that
\begin{equation}
\frac{\langle h^2 \rangle^2 }{\langle h \rangle^3} \propto \frac{c^\nu}{m^{\nu-1}}.
\label{momentratio2}
\end{equation}
Using both moment ratios (which are valid for $1<\nu <2$) in Eq. (\ref{threebis})
one obtains
\begin{equation}
f(h) = \frac {\langle h \rangle^3}{\langle h^2 \rangle^2}
F\left(\frac {h\langle h \rangle}{\langle h^2 \rangle}\right),
\label{threebisbis}
\end{equation}
which motivates the rescaling in Eqs. (\ref{rescaling1})-(\ref{rescaling2})
(that is, the scaling law (\ref{threebis}) implies the rescaling),
with the new scaling function $F$ including a power-law decay with exponent $\nu$ multiplying the original scaling function $G$.
In the case of the gamma distribution, Eq. (\ref{gamma}), 
we can include $\theta$ in the relation, i.e., 
$c\propto\theta {\langle h^2 \rangle }/{\langle h \rangle}$;
nevertheless, note that $\theta$ is a constant (in contrast to $c$, which depends on $\ell$, through the moments).
In summary, the data collapse in Fig. \ref{Dindividualcells2}(a) implies the scaling of the moments 
given by Eq. (\ref{ha2}), which will be empirically verified below. 


\subsection{Relation with the multifractal canonical partition function}

Let us see how we can relate the multifractal canonical partition function
$Z_q(\ell)$ with the moments of $h$.
The definition of $Z_q(\ell)$ is \cite{Feder,Arneodo95},
\begin{equation}
Z_q(\ell)= \sum_{i  | \mu_i > 0} \mu_i ^q,
\label{Zq}
\end{equation}
where
the sum is for the occupied cells ($\mu_i >0$),
labelled by $i$;
$\mu_i$ is the empirical probability of occupation of cell $i$ 
given by $\mu_i=h_i/M$,
with $M=\sum_i h_i$ (that is, the total population);
and
$q$ can take any value (also negative ones, as $h_i>0$).
In other words, $\mu_i$ 
is the probability that a randomly chosen individual 
(uniformly from a list of individuals) belongs to cell $i$.

As $\langle h^q\rangle$ can be calculated as 
$\sum_{i|h_i>0} h_i^q/N(\ell)$, 
with $N(\ell)$ the number of occupied cells,
it is obvious that 
\begin{equation}
Z_q(\ell)=\frac {N(\ell) \langle h^q \rangle}{M^q}.
\label{Zhq}
\end{equation}
Thus, in some sense, 
$Z_q(\ell)$ computes the moments of $f(h)$
but introducing a different normalization.
Under the multifractal scenario one has that,
over a certain $\ell-$range,
$N(\ell) \propto 1/\ell ^{d_f}$ and 
\begin{equation}
Z_q(\ell)\propto \ell^{\tau(q)},
\label{Zq2}
\end{equation}
with $d_f$ the box-counting fractal dimension
and $\tau(q)$ the so-called mass exponents,
which depend on $q$ \cite{Feder}
(in fact, $N(\ell)=Z_{q=0}(\ell)$ and thus, $\tau(0)=-d_f$).
Substituting 
$N(\ell)$ and $Z_q(\ell)$
into Eq. (\ref{Zhq}) and isolating 
\begin{equation}
\langle h^q \rangle \propto \ell^{\tau(q)+d_f}, 
\label{tl}
\end{equation}
from which one obtains
$$
\frac{\langle h^2 \rangle }{\langle h \rangle } \propto \ell^{\tau_2 },
$$
using that $\tau(1)=0$ by construction
(as $Z_{q=1}=1$) and denoting $\tau_2=\tau(2)$.
This, together with Eq. (\ref{momentratio}), allows one to establish a relation between the scaling factor $c$
in the distribution of $h$ and the cell width $\ell$, which is simply
\begin{equation}
c \propto \ell ^{\tau_2 }.
\label{altau}
\end{equation}

Now we are able to compare Eq. (\ref{tl}) with Eq. (\ref{ha2}),
arriving at
\begin{equation}
\tau(q)= -d_f \mbox{ for } q < \nu -1,
\label{tauq1}
\end{equation}
and
\begin{equation}
\tau(q)=
\tau_2 (q+1-\nu)-d_f=
\tau_2 (q-1) 
\mbox{ for } q > \nu -1,
\label{tauq2}
\end{equation}
where we have used that 
\begin{equation}
d_f=\tau_2(2-\nu),
\label{dftau2}
\end{equation}
coming from the fact that $\tau(1)=0$.
As $1 < \nu < 2$,
the previous equations include the particular values 
$\tau(0)=-d_f$ 
and
$\tau(1)=0$.
If we compute the so-called
generalized fractal dimensions \cite{Arneodo95}, 
defined as $D_q=\tau(q)/(q-1)$, these become constant ($D_q=\tau_2 $) for $q > \nu -1$ 
(notice then that the measures of diversity 
$[Z_q(\ell)]^{1/(1-q)}$, explained in Refs. \cite{Jost,Tuomisto}, 
scale as $\ell^{-D_q}$ under the multifractal scenario).

Coming back to the scaling shape of $f(h)$, Eq. (\ref{threebis}),
and substituting there Eq. (\ref{altau}) we get
\begin{equation}
f(h) = 
\frac 1 m \left(\frac {m^\nu} {\ell^{\nu\tau_2 }} \right) F\left(\frac h {\ell^{\tau_2 }}\right)=
 \frac {m^{\nu-1}} {\ell^{2\tau_2 -d_f}}  F\left(\frac h {\ell^{\tau_2 }}\right),
\label{fss}
\end{equation}
where 
we have obviated the proportionality constant between $c$ and $\ell^{\tau_2}$
and
we have also introduced the relation between $\nu$, $\tau_2 $, and $d_f$, Eq. (\ref{dftau2}).
It is clear that the previous expression for $f(h)$ constitutes a finite-size scaling law \cite{Privman,Corral_garcia_moloney_font}, 
with the cell width $\ell$ playing the role of system size; thus, in this context, the cell is the ``system''
and the region under study providing multiple copies (an ensemble) of the system.

A straightforward consequence of the scaling of moments, Eq. (\ref{tl}), is that the mean population per cell scales as $\langle h \rangle \propto \ell^{d_f}$, and the mean population density (per cell)
as $\langle \rho \rangle \propto 1/\ell^{2-d_f}$. 
The reason, is that, obviously, we do not consider unpopulated cells (this is fundamental in the two approaches that we use, the study of the distribution of $h$, and its scaling, 
and the multifractal approach). 
This has the consequence that the mean population density is an ``elusive'' concept, in the sense that it fully depends on the scale of observation.
Although this behavior arises from the fact that we do not consider empty cells, 
this cannot be considered an artifact, as it makes sense to measure the population density that the individuals see (in empty areas the individuals do not feel the emptiness).

\subsection{Empirical support}

Figure \ref{Dindividualcells2}(b) provides empirical support for the finite-size scaling law
(using the values of the parameters determined below by means of the multifractal analysis).
As the involved exponents are positive it turns out to be that in an hypothetical infinite-$\ell$ limit the exponential-like tail, given by $F$, disappears (goes to infinite) and one obtains a pure power-law tail
for the distribution of cell inhabitants, $f(h)$.
In practice, for the studied region, 
this is close to be observable for $\ell \ge 0.032$ degrees, 
see Fig. \ref{Dindividualcells}(a)
(note that although some authors refer to power-law distributions as fractal distributions, we avoid such identification, as there is no direct relation, in general; nevertheless, we present here a far-from-trivial but particular connection).


Numerical analysis fully confirms our predictions for the mass exponents. 
We can compute, for the population data, 
the partition function $Z_q(\ell)$
as a function of $\ell$ for different values of $q$,
and fit straight lines to
$\ln Z_q(\ell)$ as a function of $\ln \ell$
(one fit for each value of $q$, with the slope being $\tau(q)$).
The results appear in Table \ref{table1}.
Alternatively, we can take advantage that
the values of $\tau(q)$ are fully determined by just two parameters, 
$d_f$ and $\tau_2$, see Eqs. (\ref{tauq1}) and (\ref{tauq2}).
In this way, we perform just one fit for all values of $q<\nu-1$
to get $d_f$, and 
another single fit for $q>\nu-1$ to get $\tau_2$,
leading to 
$d_f\simeq 1.29$ and 
$\tau_2\simeq 1.67$ 
(Appendix III explains how to do these global fits).

\begin{figure}[ht]
\includegraphics[width=.90\columnwidth]{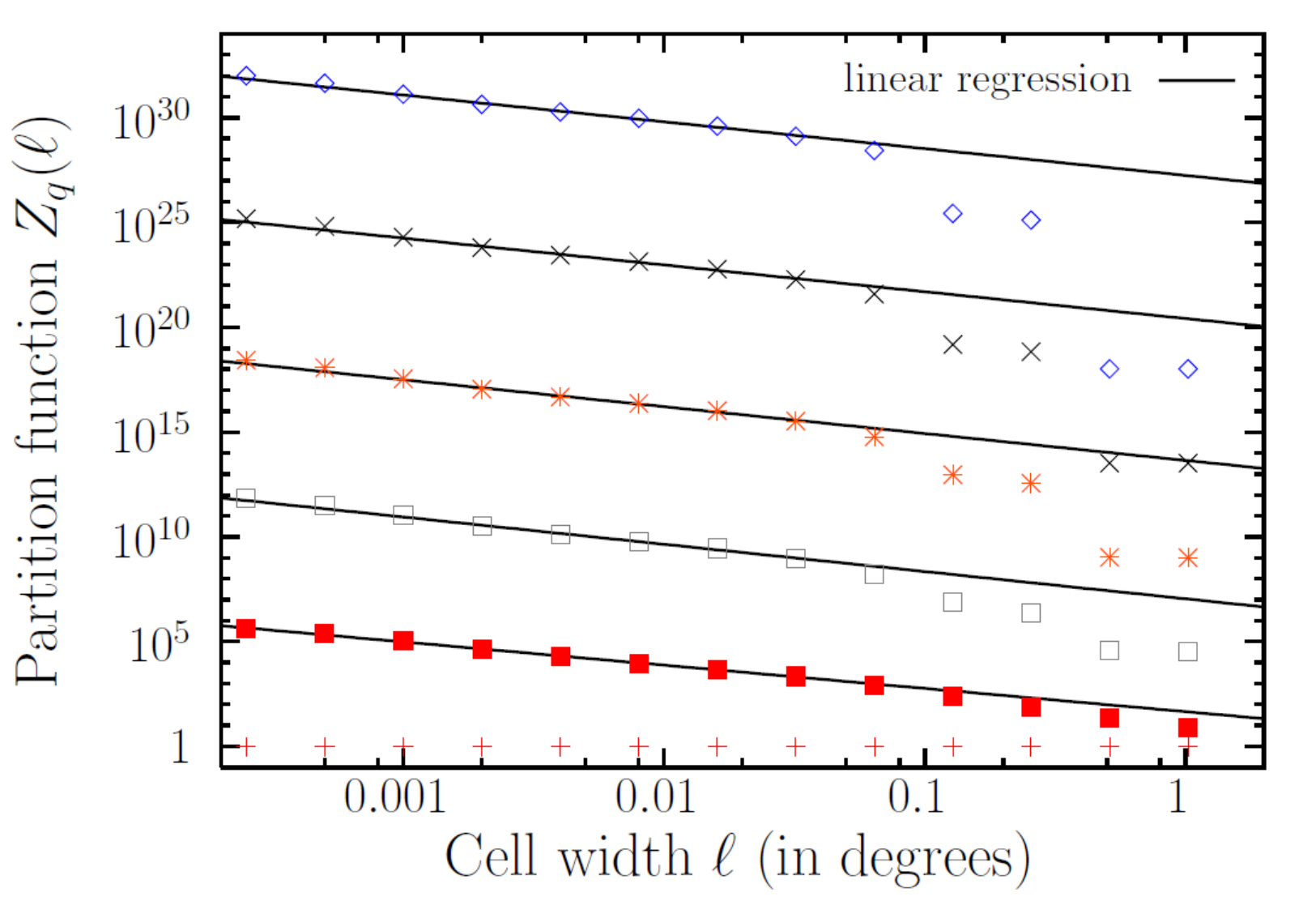}\\
\includegraphics[width=.90\columnwidth]{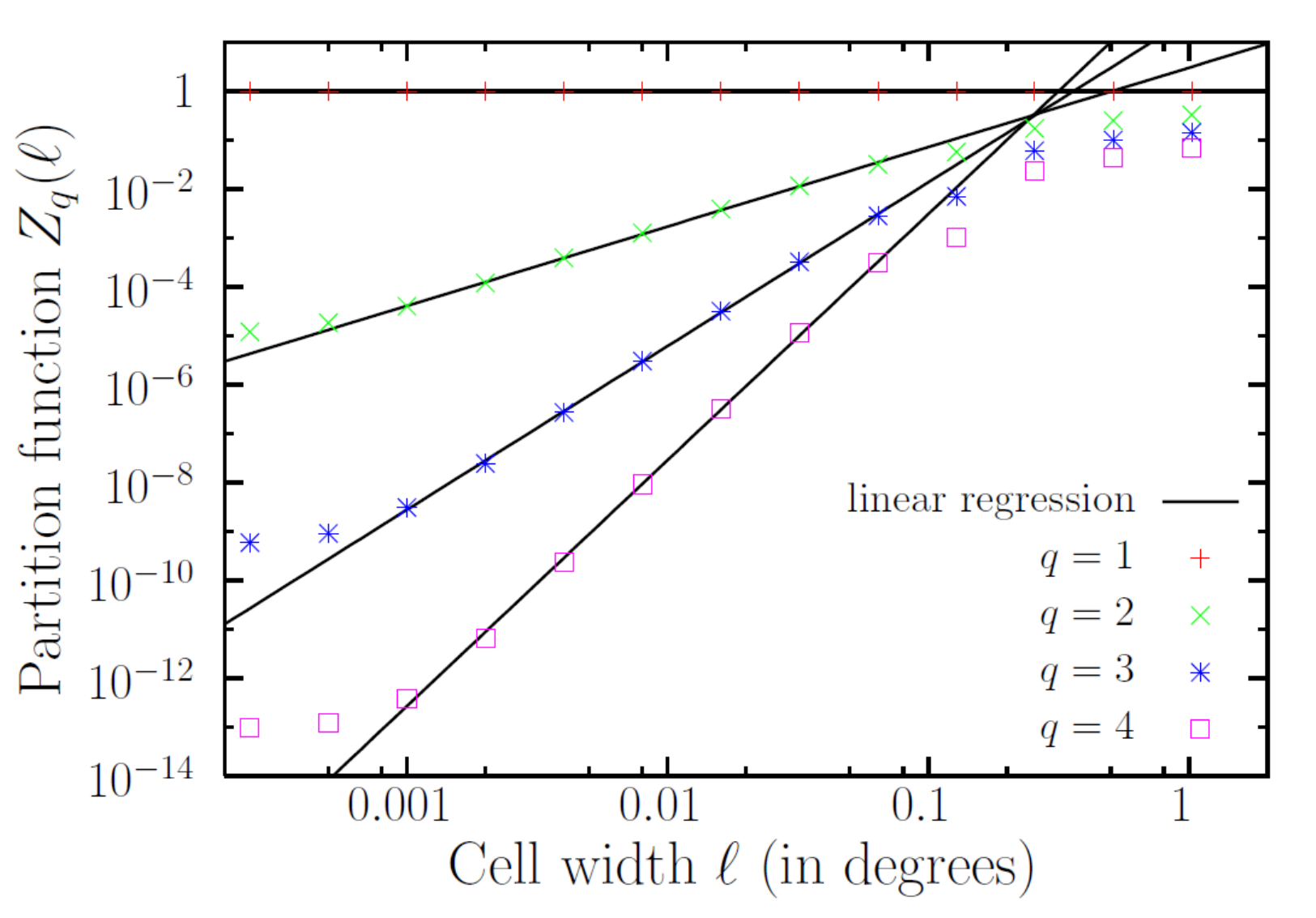}
\caption{
Power-law dependence of the partition function $Z_q(\ell)$ as a function of 
$\ell$, for different values of $q$, in agreement with the behavior predicted by Eqs. 
(\ref{tauq1}) and (\ref{tauq2}).
Values of $q$ are, from top to bottom, $q=-4$, $-3$... up to $q=4$.
The displayed linear-regression fits correspond to the individual fits whose values are provided in Table \ref{table1}.
}
\label{multif}
\end{figure}

Table \ref{table1}
shows that the results of the individual and the global fits of $Z_q(\ell)$ 
are very close to each other (for equal values of $q$),
supporting our theoretical results.
Plots with some illustrative fits are shown in Fig. \ref{multif}.
Thus, we conclude that $\tau(q)=-d_f\simeq -1.29$ for $q\le 0$
and $\tau(q)\simeq 1.67 (q-1)$  for $q\ge 1$.
From both values and Eq. (\ref{dftau2})
we can calculate the power-law exponent
$\nu=2-d_f/\tau_2\simeq 1.23$.  
Notice that the procedure of performing global fits 
is preferable to computing averages from the results of the individual fits, see Appendix III.
Notice also that this procedure allows the fitting of parameters in a model that
is not fully parametric (the scaling function is unspecified).

\begin{table}[h]
\begin{center}
\caption{\label{table1} 
Linear regression fits of $\ln Z_q(\ell)$ versus $\ln \ell$.
Two global fits are performed, 
one for $q=-4$, $-3$, $-2$, and $-1$, 
and another one for $q=2$, $3$, and $4$.
The Pearson correlation coefficient is denoted by $r$.
}
\smallskip
\newpage
\begin{tabular}{|rcccc|}
\hline
& fitting range in $\ell$ && $\tau(q)$ & $\tau(q)$\\
$q$ & (in degrees)& $r$ & indiv. fit & global fit\\
\hline 
$-4$ & 0.001 - 0.032 & $-0.997$ & $-1.279\pm0.048$ & $-1.285$ \\
$-3$ &  0.001 - 0.032 & $-0.997$ & $-1.278\pm0.049$ & id\\
$-2$ &  0.001 - 0.032 & $-0.997$ & $-1.282\pm0.051$ & id\\
$-1$ &  0.001 - 0.032 & $-0.997$ & $-1.301\pm0.050$ & id\\
$0$ &  0.001 - 0.032 &  $-0.998$ & $-1.110\pm0.031$ & --\\
\hline 
$1$& 0.001 - 0.064 & -- & $0.000$ & 0.000\\
$2$&  0.001 - 0.064 & $0.9998$ & \phantom{$-$}$1.624\pm0.016$ & 1.672\\
$3$&  0.001 - 0.064 & $0.9998$ & \phantom{$-$}$3.348\pm0.028$ & 3.344\\
$4$&  0.001 - 0.064 & $0.9996$ & \phantom{$-$}$5.030\pm0.063$ & 5.016\\
\hline 
\end{tabular}
\par
\end{center}
\end{table}

\subsection{Relation with the singularity spectrum}

Another equivalent way to characterize multifractal
behavior is by means of the multifractal spectrum, 
or singularity spectrum.
This has a much clearer geometrical interpretation than the partition function.
What one has for a multifractal is that the mass scales 
with the cell width differently from point to point \cite{Arneodo95}, 
i.e., for a point $i$ the scaling should go as 
\begin{equation}
\mu_i \propto \ell^{\alpha_i},
\label{mualpha}
\end{equation}
where $\alpha_i$ is the singularity strength at $i$
(for a monofractal $\alpha_i$ must be the same for all $i$).
%
%
%

The multifractal spectrum, or singularity spectrum \cite{Arneodo95}, $f(\alpha)$, describes
the points $i$ for which $\alpha_i$ takes a particular value $\alpha$
in such a way that $\Omega_\alpha(\ell)$ counts the number of cells
with such a value of $\alpha$, scaling as 
\begin{equation}
\Omega_\alpha(\ell) \propto \frac 1{\ell^{f(\alpha)}}
\label{omega}
\end{equation}
(note that $f(\alpha)$ is neither a probability mass function nor a probability density,
although we use the same symbol as for $f(h)$ and $f(\rho)$).
A Legendre transform relates the multifractal spectrum and the mass exponents, 
\begin{equation}
f(\alpha)= \alpha q -\tau(q), \mbox{ with }
\alpha=\frac{\partial \tau}{\partial q},
\label{multifractalspec}
\end{equation}
see our Appendix IV, or also Ref. \cite{Feder}


Applying these two equations to our case, 
we obtain, from Eqs. (\ref{tauq1}) and (\ref{tauq2}), that
\begin{equation}
\alpha = \left\{\begin{array}{ll} 
0  & \mbox{ for } q < \nu -1,\\
\tau_2  & \mbox{ for } q > \nu -1,\\
\end{array}\right.
\label{eqalpha}
\end{equation}
and
\begin{equation}
f(\alpha) = \left\{
\begin{array}{ll} 
d_f & \mbox{ for } \alpha=0,\\
\tau_2 &  \mbox{ for } \alpha=\tau_2,\\
\end{array}\right.
\label{eqfalpha}
\end{equation}
and undefined otherwise.
So, the multifractal turns out to be a bifractal.
In the analogy with statistical mechanics
we see that in the microcanonical description 
there are only two macrostates, 
with energy $0$ and $\tau_2=d_f/(2-\nu)$.

\section{Discussion and summary}

An intriguing finding of this research is that we provide an empirical realization of a new central-limit theorem (CLT).
In the usual CLT, the addition of a large but fixed number of independent identically distributed random variables (with finite variance)
leads to a normal distribution, whereas in the generalized CLT the limiting distributions are L\'evy
 distributions if the added independent variables have infinite variance \cite{Bouchaud_Georges,Mantegna_Stanley}.
Here we have an empirical distribution (fitted reasonably well by the gamma distribution) that is invariant
under addition with rescaling, not being neither normal nor L\'evy. 
So, neither the usual nor the generalized CLT is fulfilled.
The reason of this discrepancy, is, of course, the existence of strong dependence between the added variables.
To be more concrete, if we double the width of the cells
(and this $\ell^2$ transforms to $4\ell^2$), 
the resulting number of inhabitants is the sum of $4$ realizations of $h$ at the small scale ($\ell$).
The new $h$ has to be rescaled as $h/4^{\tau_2/2}=h/3.18$ to yield a ``stable'' distribution, which is a highly nontrivial result.

In summary, using high-spatial resolution data for the human occupation of a region, 
we have unveiled the existence of a finite-size scaling relation for the distribution of the number of inhabitants in fixed-size cells.
The calculation of the moments of a distribution fulfilling finite-size scaling allows to establish a relation with the multifractal partition function, and the mass exponents can be obtained from here.
The two different behaviors of the moments (depending on whether $q<\nu-1$ or $q>\nu-1$) lead to existence of only to singularity exponents, as given by Eq. (\ref{eqalpha}).
Our approach has been developed from the study of a single (small) region in Western Europe.
Given the degree of universality of this sort of phenomena, 
we expect other regions to behave very similarly, 
but, of course, this speculation has to be validated empirically.
Further, it would be of the maximum interest to study the applicability of this beyond human populations, and consider in the same framework the spatial distribution of animals and plants.

\section{Acknowledgements}

We are grateful
to E. Su\~n\'e and F. Udina
from IDESCAT, for their kindness to provide the data for a previous article \cite{Corral_Arcaute}, 
to S. Manrubia for preliminary discussions,
to S. Pueyo for providing bibliography, 
and to A. Allard and J. Serr\`a for an unfinished collaboration in a related work.
%
%
Support from projects
PGC-FIS2018-099629-B-I00
from the Spanish MICINN,
CEX2020-001084-M from
the Spanish State Research Agency,
through the Severo Ochoa and Mar\'{\i}a de Maeztu Program for Centers and Units of Excellence in R\&D,
as well as the CERCA Programme (Generalitat de Catalunya)
is acknowledged.

\section{Appendix I}

Let us start with the simplest random distribution of points in space, 
that given by a Poisson process.
If we introduce a grid of equal cells, 
the number of points (individuals or inhabitants) $h$ in each cell
will be given by the Poisson distribution, with parameter (mean value) $\lambda$.
If we double the area of the cells, the parameter $\lambda$ of the Poisson distribution 
(which is a shape parameter) doubles
($\lambda\rightarrow 2\lambda$), and the shape of the distribution changes under cell aggregation (this is a basic property of the Poisson distribution).

The use of the negative binomial distribution constitutes one step further. 
This can be justified as coming from a mixture of Poisson distributions with different parameters $\lambda$ given by the gamma distribution, i.e., 
$$
\mbox{NBD}(h;\gamma, 1/(a+1))=
\int_0^\infty \mbox{Poisson}(h;\lambda) \mbox{gamma}(\lambda;\gamma,a) d\lambda,
$$
where 
{NBD}$(h;\gamma, 1/(a+1))$ denotes a negative binomial distribution with 
shape parameters $\gamma$ and $1/(a+1)$,
{Poisson}$(h;\lambda)$ denotes a Poisson distribution with parameter (mean) $\lambda$,
and
{gamma}$(\lambda;\gamma,a)$ denotes a gamma distribution with shape $\gamma$ and scale $a$.
This means that the number of points (inhabitants) in each cell follows a Poisson distribution, 
but each cell with its own value of $\lambda$, given by the gamma distribution.
The mixture of the different Poisson distributions along all cells leads to the negative binomial.

Let us double the area of the cells in such a way that we merge contiguous cells, 
then, merging cells 1 and 2 we get, for the merged cell 
$h=h_1+h_2$, with $h_i$ Poisson distributed with parameter $\lambda_i$.
As the sum of independent Poisson is Poisson, $h$ will be Poisson distributed with parameter
$\lambda=\lambda_1+\lambda_2$.
We are interested in the distribution of $\lambda$.
Both $\lambda_1$ and $\lambda_2$ are gamma distributed, with shape parameter $\gamma$.
But the sum of gammas is gamma, with shape parameter the sum of the individual shape parameters; thus, $\lambda$ (the parameter of the Poisson distribution for the merged cells)
will be gamma distributed with shape parameter $2\gamma$.
In other words, the larger the cells, the larger the shape parameter of the gamma distribution (as it happened for the Poisson process).
This means that the resulting negative binomial distribution describing the number of points (inhabitants) in the merged cells is given by {NBD}$(h;2\gamma,1/(a+1))$, 
and the resulting distribution has a different (doubled) shape parameter.
In other words, the negative binomial, if it arises as a mixture of Poisson processes, 
cannot account for the invariance of $f(h)$ under cell aggregation observed empirically.
The negative-binomial model would be ill-defined, 
in the sense that it is only valid for the particular scale for which it is defined.


\section{Appendix II}

Let us clarify the scaling of the moments of the distribution of $h$.
For exponent $\nu>1$, the scaling of the distribution with the scale parameter $c$ is given by Eq. (\ref{threebis}), defined in the range $h>m$ with $m>0$.
Therefore, the moments of order $q$ are given by
$$
\langle h^q \rangle =\int_m^\infty h^q f(h) dh
=m^{\nu-1} \int_m^\infty dh h^{q-\nu} G(h/c)
=m^{\nu-1} c^{1+q-\nu} \int_{m/c}^\infty dz z^{q-\nu} G(z)
$$
$$
\propto m^{\nu-1} c^{1+q-\nu} 
\left[\left.\frac {z^{1+q-\nu}}{1+q-\nu}\right|_{z=m/c} + \mbox{constant}\right],
$$
where the constant does not depend neither on $c$ nor $m$.
In the limit $c\gg m$, 
$$
\langle h^q \rangle 
\propto \left\{
\begin{array}{lcr}
m^q & \mbox{ if } & 1+q-\nu<0,\\
m^{\nu-1} c^{1+q-\nu} & \mbox{ if } & 1+q-\nu>0,\\
\end{array}
\right.
$$
as shown in Eq. (\ref{ha2}).
Note that normalization requires $\nu>1$, 
as the zeroth-order moment cannot scale with $c$.

\section{Appendix III}



Let us consider $Z_q(\ell)\propto \ell^{\tau(q)}$, 
which we can write $y_q=a_q+\tau(q) x_q$, 
with $y_q=\ln Z_q(\ell)$ and $x_q= \ln \ell$
(note that the latter does not necessarily depend on $q$, 
but we keep the subindex for convenience).
In principle, if we consider $n_q$ different values of $q$,
we have to fit $2n_q$ independent parameters.
But for $q> \nu-1$ we have $\tau(q)=\tau_2 (q-1)$, 
and in this scenario we only deal with $n_q + 1$ independent parameters 
(in fact, we are not interested in the $n_q$ values of $a_q$,
and thus, we end with only one parameter of interest, $\tau_2$).
As usual, in linear least squares we have the function
$$
\sum_{\forall q\,\forall i}(y_{qi}-a_q-\tau(q)x_{qi})^2,
$$
where $x_{qi}$ and $y_{qi}$ denote the $i-$th data point 
of each respective variable $x_{q}$  and $y_{q}$.
Differentiation with respect $a_q$ and equating to zero 
leads to the usual solution $a_q=\bar y_q -\tau(q) \bar x_q$,
with $\bar x_{q}$ and $\bar y_{q}$ the respective sample means.
Differentiation with respect $\tau_2$ and substitution of $a_q$ 
leads to 
$$
\sum_{\forall q} (q-1) \left[ \mbox{cov}(x_q,y_q)
-\tau(q)s_{xq}^2\right]=0
$$
with $s_{xq}^2$ the (biased) sample variance of $x_q$,
$\mbox{cov}(x_q,y_q)$ the covariance, 
and the factor $q-1$ arising from $d\tau(q)/d \tau_2=q-1$.
Then, the solution for $\tau_2$ is
$$
\tau_2=
\frac{\sum_{\forall q} (q-1) \mbox{cov}(x_q,y_q)}
{\sum_{\forall q}(q-1)^2 s_{xq}^2},
$$
which is valid for $q>\nu-1$
(note that the case $q=1$ has no influence in the solution).
In the derivation it is implicit that all values of $q$ 
contribute with the same number of data points.

In the opposite case, corresponding to $q<\nu-1$ we
know that $\tau(q)=-d_f$, and an analogous derivation leads to
$$
d_f=
-\frac{\sum_{\forall q} \mbox{cov}(x_q,y_q)}
{\sum_{\forall q} s_{xq}^2}.
$$
The solution for $a_q$ remains the same as before.
This also holds for the relation between $\alpha$ and $f(\alpha)$
with $\ell$, with the corresponding redefinition of $y_q$.

\section{Appendix IV}

The relation for $\alpha_i$ given by Eq. (\ref{mualpha})
is not useful from an operational point of view 
(see nevertheless Ref. \cite{Turiel06}).
Instead, 
we can better use a well-known analogy with 
the ensemble theory of statistical mechanics \cite{Stauffer_stat_mech}.
The correspondence multifractals $ \longleftrightarrow$ statistical mechanics is
given by
$$
q  \, \longleftrightarrow \,  \beta, \,
$$
$$
\alpha_i   \, \longleftrightarrow \,  E_i, \,
$$
$$
\ln \frac 1 \ell  \, \longleftrightarrow \,  V, \,
$$
$$
Z_q(\ell)  \, \longleftrightarrow \,  Z_\beta, \,
$$
with
$\beta=(k_B T)^{-1}$, 
$E_i$ the energy per unit volume of a microstate $i$,
and $V$ the volume
($k_B$ is Boltzmann constant and $T$ is the temperature, and
the thermodynamic limit $V\rightarrow \infty$ corresponds to 
$\ell \rightarrow 0$).
In this way,
the multifractal partition function $Z_q(\ell)$, Eq. (\ref{Zq}), 
results exactly the same as
the partition function of the canonical ensemble in statistical mechanics,
$Z_\beta=\sum_i e^{-\beta E_i V}$,
using Eq. (\ref{mualpha}) also. 
Further, from Eq. (\ref{Zq2}),
the mass exponents turn out to be directly related to the Helmholtz free energy ${\mathcal F}$ as
$$
\tau(q)=\frac{\ln Z_q(\ell)}{\ln \ell}  \, \longleftrightarrow \,  
\frac {\beta{\mathcal F}} V = -\frac {\ln Z_\beta} V,
$$
for small $\ell$ (large $V$).

Remember that the multifractal spectrum \cite{Arneodo95}, 
under the multifractal scenario, verifies 
$\Omega_\alpha(\ell) \propto \frac 1{\ell^{f(\alpha)}}$, 
where $\Omega_\alpha(\ell)$ counts the number of cells
with such a value of $\alpha$.
Under the multifractal $ \longleftrightarrow$ statistical-mechanics correspondence,
$\Omega_\alpha(\ell)$ counts microstates with energy $V E$,
so its logarithm [Eq. (\ref{omega})] is related to the entropy through Boltzmann formula
$$
f(\alpha) = \frac {\ln \Omega_\alpha(\ell)}{\ln \ell^{-1}} 
 \, \longleftrightarrow \,  \frac S {k_B V} =\frac {\ln\Omega_E}{V},
$$
when $\ell$ is small.
Therefore, the usual Legendre transform ${\mathcal F}=U-TS$
(with $U=V E=\partial (\beta{\mathcal F})/\partial \beta$) can be written (by direct substitution)
as a Legendre transform relating the multifractal spectrum and the mass exponents \cite{Feder},
$f(\alpha)= \alpha q -\tau(q)$, 
with 
$\alpha={\partial \tau}/{\partial q}$, as stated in the main text.


%

\begin{thebibliography}{10}

\bibitem{West_book}
G.~West.
\newblock {\em Scale: The Universal Laws of Life and Death in Organisms, Cities
  and Companies}.
\newblock Penguin Press, 2017.

\bibitem{Barthelemy_cities_review}
M.~{Barthelemy}.
\newblock {The statistical physics of cities}.
\newblock {\em Nature Rev. Phys.}, 1(6):406--415, 2019.

\bibitem{Arcaute_Ramasco}
E.~Arcaute and J.~Ramasco.
\newblock Some recent advances in urban system science: models and data.
\newblock {\em arXiv}, 2110.15865, 2021.

\bibitem{Rybski_Gonzalez}
D.~Rybski and M.~C. Gonz\'alez.
\newblock Cities as complex systems--collection overview.
\newblock {\em PLOS ONE}, 17(2):1--6, 02 2022.

\bibitem{Zipf_1949}
G.~K. Zipf.
\newblock {\em Human Behavior and the Principle of Least Effort}.
\newblock Addison-Wesley, 1949.

\bibitem{Krugman}
P.~Krugman.
\newblock Confronting the mystery of urban hierarchy.
\newblock {\em J. Japan. Int. Econo.}, 10:399--418, 1996.

\bibitem{Cottineau}
C.~Cottineau.
\newblock {MetaZipf}. {A} dynamic meta-analysis of city size distributions.
\newblock {\em PLOS ONE}, 12(8):e0183919, 2017.

\bibitem{footnote_corral_garcia_del_muro2}
Reference \cite{Corral_Cancho} explains alternative representations of {Zipf's}
  law.

\bibitem{Corral_Cancho}
A.~Corral, I.~Serra, and R.~{Ferrer-i-Cancho}.
\newblock Distinct flavors of {Zipf's} law and its maximum likelihood fitting:
  Rank-size and size-distribution representations.
\newblock {\em Phys. Rev. E}, 102:052113, 2020.

\bibitem{Eeckhout}
J.~Eeckhout.
\newblock Gibrat's law for (all) cities.
\newblock {\em Amer. Econ. Rev.}, 94(5):1429--1451, 2004.

\bibitem{Levy_comment}
M.~Levy.
\newblock Gibrat's law for (all) cities: Comment.
\newblock {\em Amer. Econ. Rev.}, 99(4):1672--1675, 2009.

\bibitem{Malevergne_Sornette_umpu}
Y.~Malevergne, V.~Pisarenko, and D.~Sornette.
\newblock Testing the {Pareto} against the lognormal distributions with the
  uniformly most powerful unbiased test applied to the distribution of cities.
\newblock {\em Phys. Rev. E}, 83:036111, 2011.

\bibitem{Corral_Arcaute}
A.~Corral, F.~Udina, and E.~Arcaute.
\newblock Truncated lognormal distributions and scaling in the size of
  naturally defined population clusters.
\newblock {\em Phys. Rev. E}, 101:042312, 2020.

\bibitem{Rozenfeld}
H.~D. Rozenfeld, D.~Rybski, J.~S. Andrade, M.~Batty, H.~E. Stanley, and H.~A.
  Makse.
\newblock Laws of population growth.
\newblock {\em Proc. Natl. Acad. Sci. USA}, 105(48):18702--18707, 2008.

\bibitem{Jiang2011}
B.~Jiang and T.~Jia.
\newblock Zipf's law for all the natural cities in the {United States}: a
  geospatial perspective.
\newblock {\em Int. J. Geograp. Inform. Sci.}, 25(8):1260--1281, 2011.

\bibitem{footnote_corral_garcia_del_muro3}
Comparisons between lognormal and power laws in other contexts are performed in
  {Refs.} \cite{Corral_Gonzalez,Serra_Corral_Zipf}.

\bibitem{Corral_Gonzalez}
A.~Corral and A.~Gonz\'alez.
\newblock Power law distributions in geoscience revisited.
\newblock {\em Earth Space Sci.}, 6(5):673--697, 2019.

\bibitem{Serra_Corral_Zipf}
M.~Serra-Peralta, J.~Serr\`a, and A.~Corral.
\newblock Lognormals, power laws and double power laws in the distribution of
  frequencies of harmonic codewords from classical music.
\newblock {\em Sci. Rep.}, 12:2615, 2022.

\bibitem{Borregard}
M.~K. Borregaard, D.~K. Hendrichsen, and G.~Nachman.
\newblock Spatial distribution.
\newblock {\em Encyclopedia of Ecology}, pages 3304--3310.

\bibitem{Zillio}
T.~Zillio and F.~He.
\newblock Modeling spatial aggregation of finite populations.
\newblock {\em Ecology}, 91(12):3698--3706, 2010.

\bibitem{Rae}
A.~Rae.
\newblock Think your country is crowded? {These} maps reveal the truth about
  population density across {Europe}.
\newblock {\em The Conversation}, January
  23:https://theconversation.com/think--your--country--is--crowded--these--maps--reveal--the--truth--about--population--density--across--europe--90345,
  2018.

\bibitem{Sune}
E.~S. Luis.
\newblock Hacia un registro estad\'{\i}stico de territorio.
\newblock {\em XXXV Congreso Nacional de Estad\'{\i}stica e Investigaci\'on
  Operativa, IX Jornadas de Estad\'{\i}stica P\'ublica}, 2015 (in Spanish).

\bibitem{Orozco}
C.~D.~V. Orozco, J.~Golay, and M.~Kanevski.
\newblock Multifractal portrayal of the {Swiss} population.
\newblock {\em arXiv}, 1308.4038, 2013.

\bibitem{Semecurbe}
F.~S\'em\'ecurbe, C.~Tannier, and S.~G. Roux.
\newblock Spatial distribution of human population in {France}: Exploring the
  modifiable areal unit problem using multifractal analysis.
\newblock {\em Geograp. Anal.}, 48(3):292--313, 2016.

\bibitem{footnote_corral_garcia_del_muro}
It would have been more precise to remove boundary cells, as these are not
  expected to be ``complete'' by construction, nevertheless, as the number of
  cells is very large, the distortion in population counts coming from boundary
  cells can be considered small.

\bibitem{Corral_csf}
A.~Corral.
\newblock Scaling in the timing of extreme events.
\newblock {\em Chaos. Solit. Fract.}, 74:99--112, 2015.

\bibitem{Christensen_Moloney}
K.~Christensen and N.~R. Moloney.
\newblock {\em Complexity and Criticality}.
\newblock Imperial College Press, London, 2005.

\bibitem{Peters_Deluca}
O.~Peters, A.~Deluca, A.~Corral, J.~D. Neelin, and C.~E. Holloway.
\newblock Universality of rain event size distributions.
\newblock {\em J. Stat. Mech.}, P11030, 2010.

\bibitem{Feder}
J.~Feder.
\newblock {\em Fractals}.
\newblock Plenum Press, New York, 1988.

\bibitem{Arneodo95}
A.~Arneodo, E.~Bacry, and J.F. Muzy.
\newblock The thermodynamics of fractals revisited with wavelets.
\newblock {\em Physica A}, 213(1):232 -- 275, 1995.

\bibitem{Jost}
L.~Jost.
\newblock Entropy and diversity.
\newblock {\em Oikos}, 113(2):363--375, 2006.

\bibitem{Tuomisto}
H.~Tuomisto.
\newblock A consistent terminology for quantifying species diversity? {Yes}, it
  does exist.
\newblock {\em Oecologia}, 164(4):853--860, 2010.

\bibitem{Privman}
V.~Privman.
\newblock Finite-size scaling theory.
\newblock In V.~Privman, editor, {\em Finite Size Scaling and Numerical
  Simulation of Statistical Systems}, pages 1--98. World Scientific, Singapore,
  1990.

\bibitem{Corral_garcia_moloney_font}
A.~Corral, R.~Garcia-Millan, N.~R. Moloney, and F.~Font-Clos.
\newblock Phase transition, scaling of moments, and order-parameter
  distributions in {Brownian} particles and branching processes with
  finite-size effects.
\newblock {\em Phys. Rev. E}, 97:062156, 2018.

\bibitem{Bouchaud_Georges}
J.-P. Bouchaud and A.~Georges.
\newblock Anomalous diffusion in disordered media: statistical mechanisms,
  models and physical applications.
\newblock {\em Phys. Rep.}, 195:127--293, 1990.

\bibitem{Mantegna_Stanley}
R. N. Mantegna and H. E. Stanley.
\newblock An Introduction to Econophysics.
\newblock Cambridge Univ. Press, Cambridge, UK, 2000.

\bibitem{Turiel06}
A.~Turiel, C.~J. P\'erez-Vicente, and J.~Grazzini.
\newblock Numerical methods for the estimation of multifractal singularity
  spectra on sampled data: A comparative study.
\newblock {\em J. Comp. Phys.}, 216(1):362--390, 2006.

\bibitem{Stauffer_stat_mech}
D.~Chowdhury and D.~Stauffer.
\newblock {\em Principles of Equilibrium Statistical Mechanics}.
\newblock John Wiley \& Sons, Ltd. Weinheim, 2000.

\end{thebibliography}


\end{document}